\documentclass[twocolumn]{aastex6}
\turnoffedit

\shortauthors{Wakita et al.}

\begin{document}
\title{Planetesimal collisions as a chondrule forming event}

\author{Shigeru Wakita\altaffilmark{1}, Yuji Matsumoto\altaffilmark{1,2}, Shoichi Oshino\altaffilmark{1}, and Yasuhiro Hasegawa\altaffilmark{3}}
\altaffiltext{1}{Center for Computational Astrophysics, National Astronomical Observatory of Japan, Mitaka, Tokyo 181-8588, Japan; shigeru@cfca.jp}
\altaffiltext{2}{Planetary Exploration Research Center, Narashino, Chiba 275-0016, Japan}
\altaffiltext{3}{Jet Propulsion Laboratory, California Institute of Technology, Pasadena, CA 91109, USA}

\begin{abstract}
Chondritic meteorites contain unique spherical materials named chondrules: 
sub-mm sized silicate grains once melted in a high temperature condition in the solar nebula. 
We numerically explore one of chondrule forming processes, planetesimal collisions. 
Previous studies found that impact jetting via protoplanet-planetesimal collisions make chondrules 
with an amount of 1 \% of impactors' mass, when impact velocity exceeds 2.5 km s$^{-1}$. 
Based on the mineralogical data of chondrules, 
undifferentiated planetesimals would be more suitable for chondrule-forming collisions than potentially differentiated protoplanets.
We examine planetesimal-planetesimal collisions using a shock physics code and find two things: 
one is that planetesimal-planetesimal collisions produce the nearly same amount of chondrules as protoplanet-planetesimal collisions ($\sim$ 1 \%).
The other is that the amount of produced chondrules becomes larger as the impact velocity increases 
when two planetesimals collide with each other.
We also find that progenitors of chondrules can be ejected from deeper regions of large targets (planetesimals or protoplanets) than small impactors (planetesimals).
The composition of targets is therefore important to fully account for the mineralogical data of currently sampled chondrules.
\end{abstract}

\keywords{meteorites, meteors, meteoroids -- planets and satellites: formation -- minor planets, asteroids: general}

\section{Introduction} \label{intro}
Chondrules are sub-mm sized spherical materials found in chondritic meteorites. 
Their main component is olivine and pyroxene and they are up to 80 \% in volume in ordinary chondrites 
and 20 \% in carbonaceous chondrites \citep{sk05,s07}. 
Precursors of chondrules experienced flash heating in the solar nebula at or above the temperature of $\sim$ 1800K,
which is the melting temperature of silicate.
Melted droplets cooled down under the existence of gas in the solar nebula with a cooling rate of about 10 - 1000 K/hour \citep{hcl05}. 
The cooling rate is required to explain chondrule textures.  
Isotope chronology of chondrules (e.g., $^{206}$Pb-$^{207}$Pb and $^{26}$Al-$^{26}$Mg) revealed that 
chondrule formation started at the time of Ca-Al rich inclusions (CAIs) formation, 
which occurred at 4567 Myr ago, and lasted for 3 - 5 Myr \citep{cbk12, bcb15}.  

Several scenarios are proposed for forming chondrules \citep{dac14}. 
Planetesimal collisions are one of the promising candidates to form chondrules \citep{ss12,jmm15,hwmo16}.
This is partly because a magnetized record obtained in the Semarkona ordinary chondrite supports a collisional origin to form chondrules \citep{fwl14}, 
and partly because unique features and Pb-Pb chondrules dating in CB chondrites also support the collisional origin \citep{kacm05,bcb15}. 
Impact jetting which can occur during planetesimal collisions could produce chondrules, 
when the impact velocity of planetesimals ($v_{imp}$) exceeds 2.5 km s$^{-1}$ ($v_{c}$), 
as shown by a shock physics code \citep{jmm15}. 
The previous study suggests that the fractional mass of planetesimals producing chondrules via impact jetting ($F_{ch}$) is about 1 \%.  
Both a Monte Carlo approach and a semi-analytical model find that 
the resultant abundance of chondrules formed via impact jetting during the formation of a protoplanet is high enough 
to account for the present asteroid belt mass \citep{jmm15,hwmo16}.
 
Previous studies implicitly assume that 
all kinds of planetesimal collisions could be modelled by collisions of a spherical body with a flat plate \citep{jmm15,hwmo16}. 
This assumption is reasonable for collisions between big protoplanets and small planetesimals, 
which can occur at a planetary accretion stage such as the oligarchic growth stage. 
However, if a protoplanet had already differentiated at the stage
\footnote{As pointed out by \citet{jmm15}, there may still be a possibility that protoplanets do not fully experience differentiation especially at their surface regions if the protoplanets grow with time (see \S\ref{sec:diff} for a more complete discussion).}, 
the composition of ejecta might be different from silicate.
In other words, the composition of resultant chondrules may not be fully consistent with the mineralogy of chondrules found in chondrites 
and hence such chondrules may not be entirely useful to reproduce the currently sampled chondrules. 
Additionally, planetesimal-planetesimal collisions are frequent at the runaway growth stage of planetary accretion, 
which is suggested by N-body simulations (S. Oshino et al. 2016 in preparation). 
This means that planetesimal-planetesimal collisions could contribute to chondrule formation for a longer time than protoplanet-planetesimal collisions
(since the latter ones become effective only after protoplanets form).
Thus, it is very important to investigate planetesimal-planetesimal collisions as a chondrule-forming event.
More specifically, it is essential to model planetesimal-planetesimal collisions 
by collisions between spherical bodies (planetesimal-planetesimal) rather than spherical-flat ones (planetesimal-protoplanet).  

In this paper, we consider various kinds of collisions and compute the ejected mass via impact jetting. 
Especially, planetesimal-planetesimal collisions are investigated using the iSALE-2D shock physics code 
to obtain a better understanding of how useful impact jetting is to form chondrules. 

\section{Methods} \label{mod}

We perform numerical simulations of planetesimal collisions 
to evaluate the resultant mass of ejecta using a shock physics code, iSALE-2D \citep{wcm06}, 
the version of which is iSALE-Chicxulub. 
The iSALE is developed to model and understand impacts and cratering, 
based on the SALE hydrocode \citep{arh80}. 
The code has been improved from SALE 
by including various equations of state (EOS), strength model, and a porosity compaction model \citep{mra92, idn97, cmi04, wcm06, cmw11}.  

Compositions of planetesimals would be related to their formation process in the solar nebula.
Formation of planetesimals is still controversial.
One of the promising scenarios is the streaming instability \citep{yg05}. 
In this scenario, planetesimals can be formed in an order of the Keplerian time at each orbit \citep[e.g.,][]{jom07}. 
If this would be the case, planetesimals next to each other should have formed at the almost same time. 
Then it can be expected that they would experience a similar degree of thermal evolution 
and their composition would also be similar  \citep[see review][]{gtbs14}. 
We assume that both of target and impactor planetesimals have the same composition in our simulations. 
Since chondrules are composed of silicates, precursors of them should be similar to silicate in their composition. 
Ordinary chondrites have chondrules 80 \% in volume and most of them are free of water. 
Bulk composition of ordinary chondrites is therefore the nearly same as that of chondrules.
On the other hand, carbonaceous chondrites contain carbon, water and less chondrules than ordinary chondrites (20 \%), 
so their bulk composition is different from chondrules.
In our simulations, we choose dunite for a bulk material of planetesimals,
such that they are similar to ordinary chondrites \citep{ss15}. 
In the actual computations, we use ANEOS equation of state for dunite. 
ANEOS is a semi-analytical model derived from the first principles of thermodynamics.
Using ANEOS,  thermodynamical quantities of materials (such as temperature, pressure and density) are obtained self-consistently \citep{tl72,m07}.
We also assume in our simulations that an initial temperature of planetesimals is uniform across their whole volume and its value is set as 300 K, 
which is similar to the gas temperature at $\sim$ 1 au in the solar nebula \citep{h81}. 
Porosity of planetesimal is fixed at 1$\%$ in this paper. 
Other input parameters for iSALE simulations are summarized in Table \ref{tab:param}; we follow \citet{jmm15} to select certain values for most of the parameters. 

\begin{deluxetable}{ll}
\tablecaption{iSALE input parameters \label{tab:param}}
\tablenum{1}
\tablehead{
\colhead{Description} & \colhead{Value}
}
\startdata
Diameter of impactor (projectile) & 10 km\\
Diameter of target & 10 - 50 km\\
Cell per projectile radius & 400 \\
Cell size (fiducial setup)  & 12.5 m \\
Initial temperature of impactor/target & 300 K \\
Bulk material of impactor/target & dunite \\
Equation of state & ANEOS \\
Solidus temperature & 1373 K \\
Simon approximation constant A \tablenotemark{a} & 1520 Mpa \\
Simon approximation exponent C \tablenotemark{a} & 4.05 \\
Poisson's ratio \tablenotemark{a} & 0.25 \\
Thermal softening parameter \tablenotemark{a} & 1.1 \\
Strength model \tablenotemark{b} & Rock \\
Cohesion (damaged) \tablenotemark{a} & 0.01 MPa \\
Cohesion (undamaged) \tablenotemark{a} & 5.07 MPa \\
Frictional coefficient (damaged) \tablenotemark{a} & 0.63 \\
Frictional coefficient (undamaged) \tablenotemark{a} & 1.58 \\
Strength at infinite pressure \tablenotemark{a} & 3.26 GPa \\
Damage model \tablenotemark{c} & Ivanov \\
Minimum failure strain &  $10^{-4}$ \\
Damage model constan &  $10^{-11}$ \\
Threshold pressure for damage model &  300 MPa \\
Porosity model \tablenotemark{d} & Wunnema \\ 
Porosity & 0.01 \\
Rate of porous compaction \tablenotemark{a} & 0.98 \\
\enddata
\tablenotetext{a}{\citet{jmm15}}
\tablenotetext{b}{\citet{idn97}}
\tablenotetext{c}{\citet{cmi04}}
\tablenotetext{d}{\citet{wcm06}}
\end{deluxetable}

There are two important parameters in this study: 
an impact velocity of planetesimals and the diameter of target planetesimals. 
The impact velocity of planetesimals ($v_{imp}$) varies from 1.0 km s$^{-1}$ to 4.0 km s$^{-1}$.
Note that the previous work finds that $v_{imp}$ of 2.5 km s$^{-1}$ is a thershold velocity to produce chondrules \citep{jmm15}.
The above range of the impact velocities can surely be realized during planetary accretion \citep{jmm15,hwmo16,htm16}. 
The diameter of impactor planetesimals is fixed at 10 km and that of target planetesimals varies from 10 km to 50 km. 
We also perform a simulation in which a flat target is considered, in order to compare with the previous work.
We adopt a cell per projectile radius (CPPR) of 400 in most of our simulations. Equivalently, the spatial cell size is 12.5 m.
When we increase the size of targets, the value of CPPR and the cell size remain the same, 
but the total number of cells increases in the simulations. 
We will verify below how the results are independent of our choice of CPPR (see \S\ref{sec:fiducial} and Figure \ref{fig:cppr}).

We define ejecta from planetesimal collisions as progenitors of chondrules 
only when a temperature of the ejecta exceeds the melting temperature ($T_m$) and 
their velocity is also larger than the impact velocity of planetesimals ($v_{imp}$) at the same time.
In order to compute $T_{m}$ realistically, 
we make use of the so-called Simon equation rather than relying on the results of ANEOS.
This is because ANEOS tends to overestimate the temperature of materials 
when its value exceeds their solidus temperature;
ANEOS uses a low heat capacity for materials in a liquid phase \citep[][]{kss12}. 
In other words, the latent heat to melt materials is not included in ANEOS.
In the Simon equation, $T_m$ is given as \citep[e.g.,][]{wco08},
\begin{eqnarray}
	T_m &=& T_0 \left( \frac{P}{A} + 1 \right)^{1/C} \label{eq:sim}, 
\end{eqnarray}
where $A$ and $C$ are Simon parameters (see Table \ref{tab:param}).
We thus separately calculate the melting temperature of ejecta as a function of pressure ($P$) in a post-processing fashion.
Based on our preliminary results, 
the total mass of chondrules estimated from this formula is about a half of the value 
that is derived from a constant solidus temperature of $T_m=1373$ K.
Since usage of the pressure-dependent $T_m$ gives a lower estimate, 
which is more conservative, we adopt this procedure in this paper. 

It should also be noted that the impact velocity ($v_{imp}$) is used to define melting ejecta as progenitor of chondrules in the above criteria.
In general, the escape velocity of the system can be utilized as the critical velocity of ejecta
since the velocity regulates the fate of ejecta.
In this study, however, the main targets are planetesimal-planetesimal collisions.
For this case, the escape velocity of the system becomes much lower than the impact velocity,
which is about a few km s$^{-1}$.
Even when protoplanet-planetesimal collisions are considered,
the escape velocity is comparable to the impact velocity.
Furthermore, our preliminary results show that 
the total amount of chondrules formed via impact jetting does not change very much
even if either the escape velocity or the impact velocity is used as a critical velocity for defining melting ejecta as progenitor of chondrules.
This is simply because the melting temperature ($T_m$) serves as a more crucial criterion than the velocity of ejecta.
We thus choose the impact velocity as a critical velocity for regarding ejecta as potential chondrules.

Based on the above numerical simulations, 
we will below discuss from which regions and what amount of chondrules are formed via impact jetting.
In order to explicitly quantify the produced amount of chondrules, 
we use three quantities ($F_{ch}$, $df_{ch}$, and $f_{ch}$).
The value of $F_{ch}$ represents what fraction of impactors' mass is transformed to chondrules.
In other words, the total chondrule mass formed via single collisions is labeled as $F_{ch} m_{pl}$, where $m_{pl}$ is the mass of impactors.
We also attempt to specify the original locations at which progenitors of chondrules are ejected,
and to quantify their mass.
For this purpose, we define $df_{ch}(r)$ as a mass fraction of impactors that is ejected at a depth ($r$) from the surface of the impactors/targets 
that eventually becomes chondrules. 
Finally, we will compute the cumulative mass fraction ($f_{ch}(r)$) of ejecta, 
which is integrated from the center of impactors/targets to a certain depth ($r$) from the surface:
\begin{equation}
\label{eq:f_ch_r}
	f_{ch}(r) = \int_{center}^{r} df_{ch}(r). 
\end{equation}
Note that at the location of the surface ($r=0$),
\begin{equation}
\label{eq:F_ch}
	f_{ch}(r=0) \equiv F_{ch}. 
\end{equation}
In the following sections, we examine how these three quantities ($F_{ch}$, $df_{ch}$, and $f_{ch}$) behave 
as a function both of the impact velocity of planetesimals and of the diameter of target planetesimals.

\section{Results} \label{res}

\begin{figure}
\figurenum{1}
\gridline{\fig{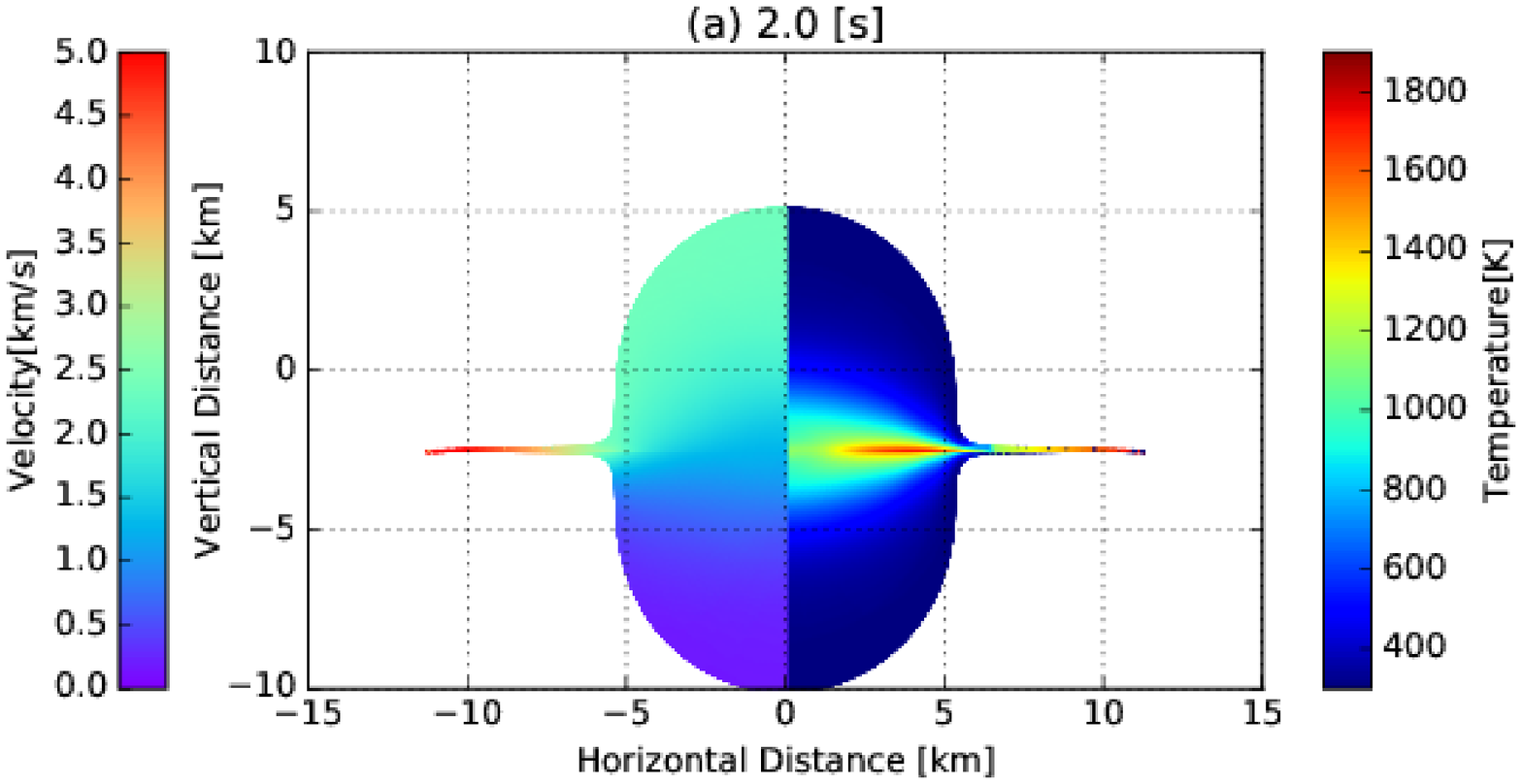}{0.5\textwidth}{}}
\gridline{\fig{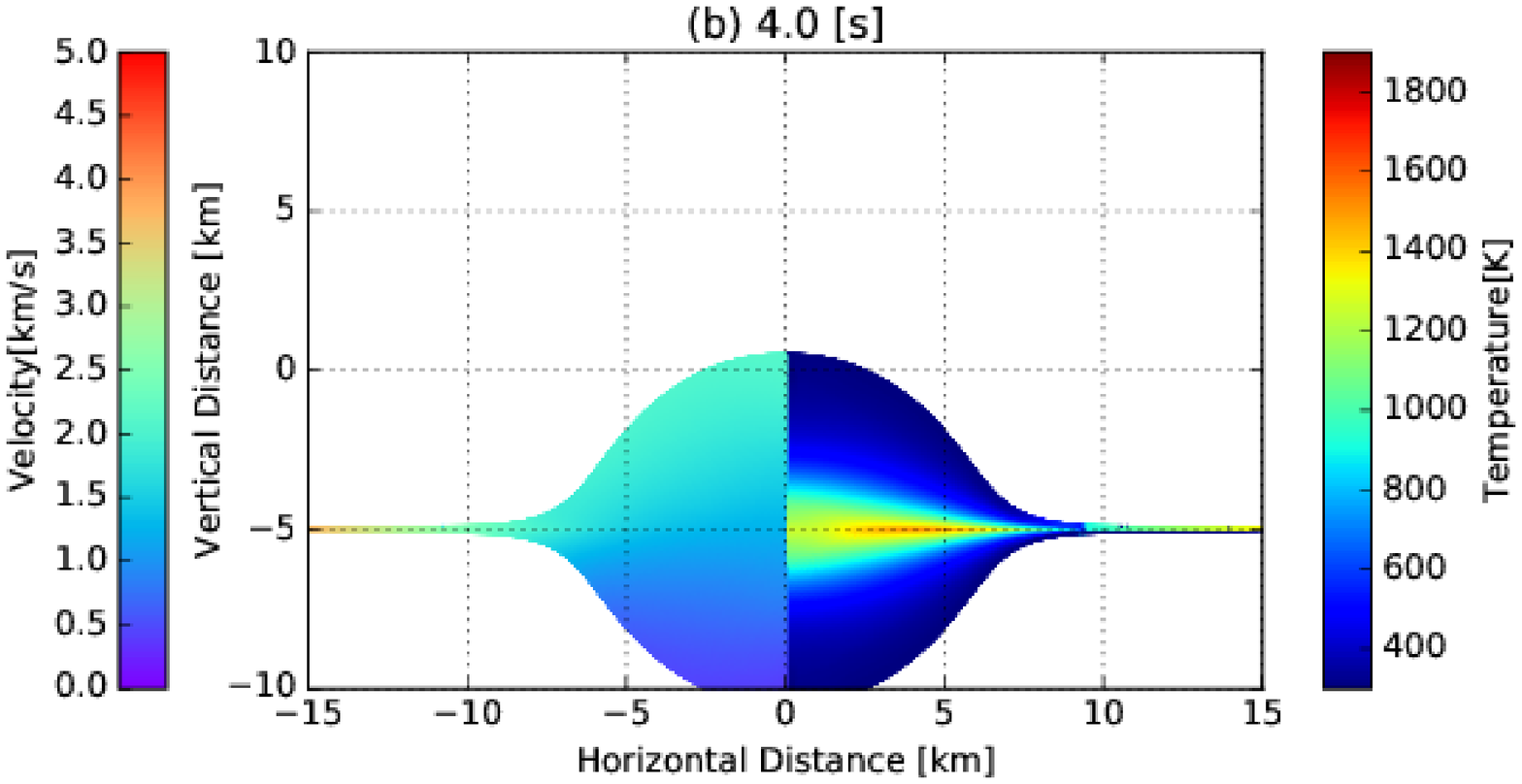}{0.5\textwidth}{}}
\caption{Snapshots of a collision between 10 km sized planetesimals with $v_{imp}=$ 2.5 km s$^{-1}$ at 2.0  s (top) and 4.0 s (bottom) after an impact.
Velocities and temperatures of materials at each position are shown on the left and right sides with the corresponding color bar, respectively.
\label{fig1}}
\end{figure}

\begin{figure}
\figurenum{2}
\gridline{\fig{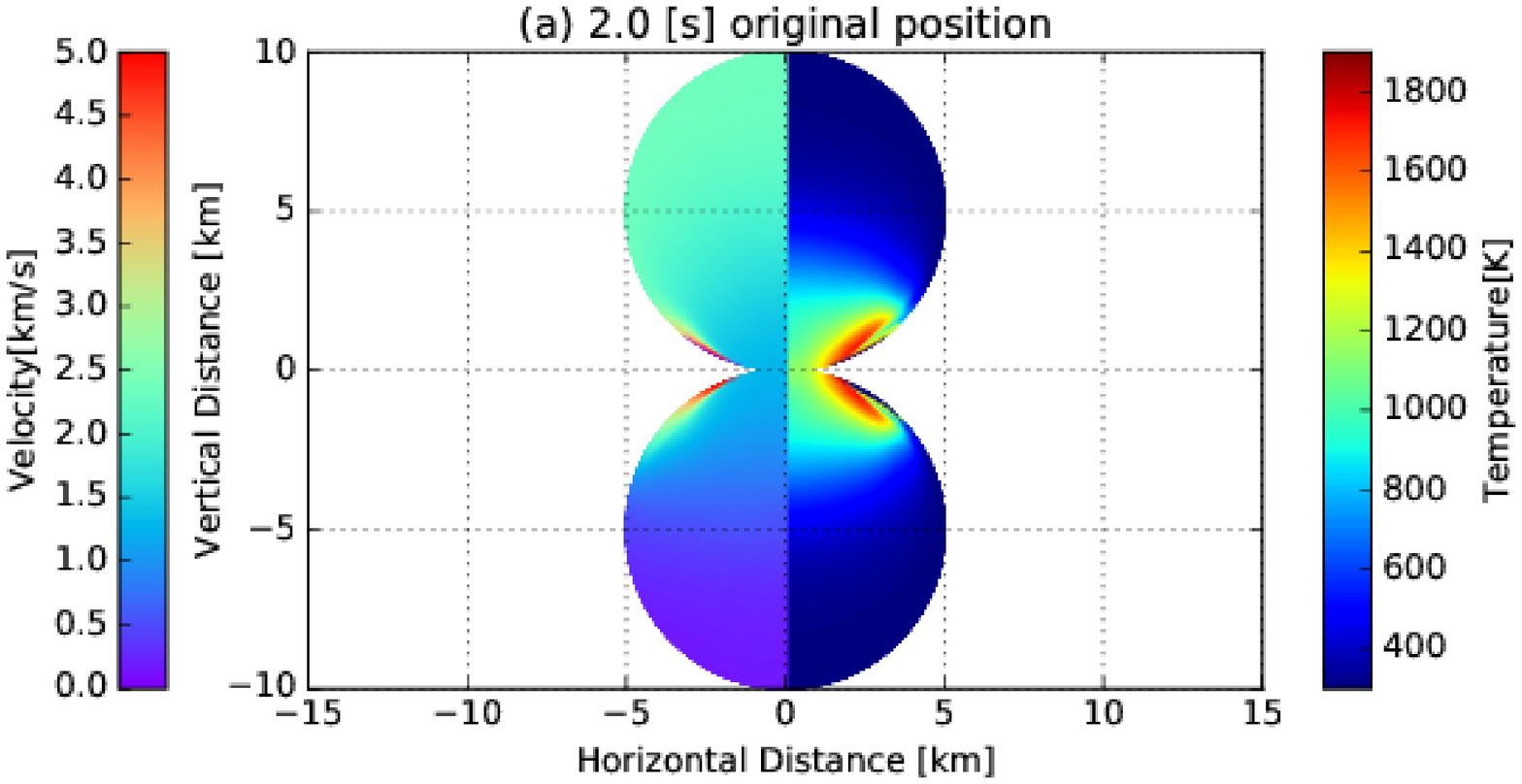}{0.5\textwidth}{}}
\gridline{\fig{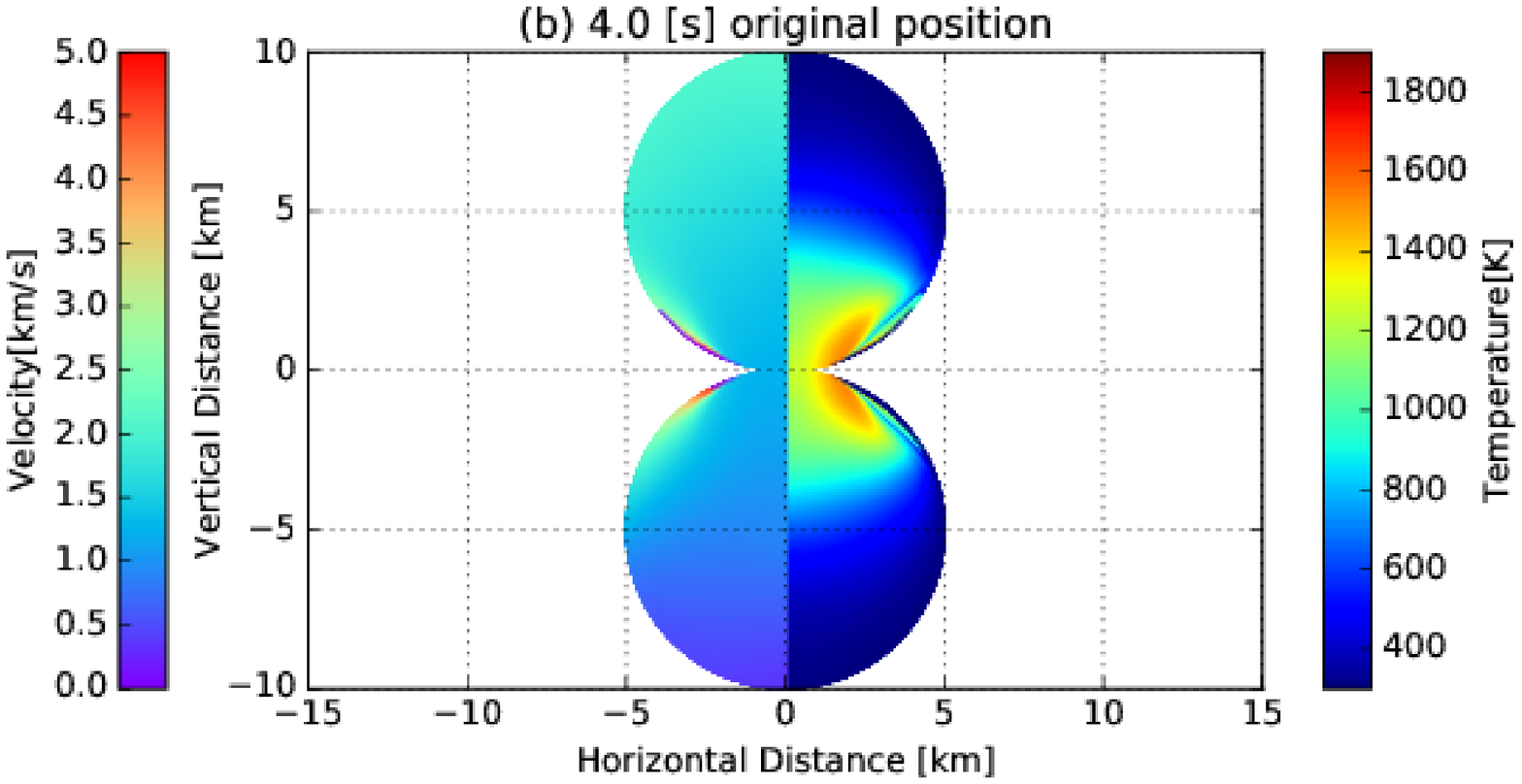}{0.5\textwidth}{}}
\caption{Snapshots of a collision between 10 km sized planetesimals with $v_{imp}=$ 2.5 km s$^{-1}$ at 2.0 s (top) and 4.0 s (bottom) after an impact (as in Figure \ref{fig1}).
In order to identify from which regions progenitor of chondrules are ejected, we trace back to their original location.
\label{fig2}}
\end{figure}

\begin{figure}
\figurenum{3}
\plotone{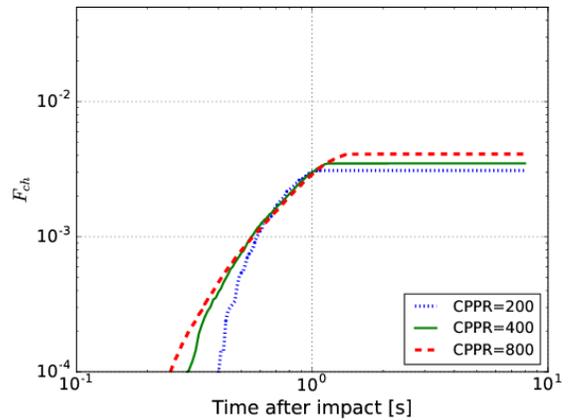}
\caption{The resultant value of $F_{ch}$ as a function of time after an impact (see equation (\ref{eq:F_ch})).
Each line represents different CPPR; 
three of them are the results obtained from collisions of a 10 km sized target with an impactor of the same size.
The impact velocity of all the collisions is 2.5 km s$^{-1}$.
\label{fig:cppr}}
\end{figure}

\begin{deluxetable}{llll}
\tablecaption{Parameters for simulation runs \label{tab:run}}
\tablenum{2}
\tablehead{
\colhead{Section} & \colhead{Diameter of target} & \colhead{Impact velocity($v_{imp}$)}
}
\startdata
\S\ref{sec:fiducial}  & 10 km & 2.5  km s$^{-1}$ \\
\S\ref{sec:10km}  & 10 km & 1.0 - 4.0  km s$^{-1}$ \\
\S\ref{sec:v40}  & 10 km, 20 km, 40 km, flat & 4.0 km s$^{-1}$ \\
\S\ref{sec:all}  & 10 km, 20 km, 40 km, flat  & 1.0 - 4.0 km s$^{-1}$ \\
\enddata
\end{deluxetable}

We present the results that are obtained from the iSALE simulations.
Various kinds of collisions are considered,
and the corresponding sets of parameters used in the simulations are summarized in Table \ref{tab:run}.

\subsection{Fiducial model: 10km sized target with impact velocity of $2.5 km s^{-1}$} \label{sec:fiducial}

Figure \ref{fig1} shows snapshots of a collision between the same sized planetesimals (10 km in diameter).
Each panel denotes temperatures on the right-side and velocities on the left-side at each position at a time of 2.0 s and 4.0 s after the impact.
We find that narrow flow (so called jet) emerges from the collisional surface between the planetesimals.
Some fraction of such ejecta could eventually become chondrules
because both of their temperatures and velocities exceed $T_m$ and $v_{imp}$, respectively.
As time goes on, the temperature of ejecta decreases (see Figure \ref{fig1} (b)).
This indicates that materials ejected at a later time are very unlikely to become chondrules
since they cannot reach $T_m$ (see below and Figure \ref{fig:cppr}).

Figure \ref{fig2} also shows the same physical quantities,
but the original positions of ejects are identified in order to specify from which regions of planetesimals the ejected materials emerge.
As seen from Figure \ref{fig2}, 
the progenitor of chondrules originates  from the surface region of the impactor and the target.
In addition, our results show that 
the temperature and velocity of materials located in the central region of the planetesimals 
are not affected by the collision very much.
In other words, ejecta from such a region cannot contribute to chondrule formation.
We will confirm below that these are the general trend for various kinds of collisions (see \S\ref{sec:10km} and Figure \ref{fig:10kmDepth}). 

It is important to quantitatively examine how much mass of chondrules are formed by single collisions as a function of time.
Figure \ref{fig:cppr} shows the time evolution of $F_{ch}$ (see the green solid line, also see equation (\ref{eq:F_ch})).
We find that the value of $F_{ch}$ initially increases with time. 
It then becomes constant at a time of $\sim$ 2.0 s or later after the impact. 
Our simulations show that the total mass of ejecta increases after 2.0 s.
These ejecta, however, do not end up with chondrules.
This is simply because ejecta which emerges at a later time do not experience melting at that time (see Figure \ref{fig1}). 
Thus, $F_{ch}$ takes a similar value between 2.0 s and 4.0 s.
We also confirm that the resultant value of $F_{ch}$ does not change after 4.0 s.
Furthermore, our results show that the total mass of chondrules' progenitor ($F_{ch}$) both from an impactor and from a target are similar (see \S\ref{sec:10km} and Figure \ref{fig:10kmDepth}).
In the following sections, we will use the value of $F_{ch}$ at 2.0 s after the impact.

We now check the convergency of our results by changing the value of cell per projectile radius (CPPR),
that is, the cell size. 
Figure \ref{fig:cppr} shows the value of $F_{ch}$ for different CPPR with the same impact velocity (2.5 km s$^{-1}$).
Each line represents different CPPR.
Our results show that there is no significant difference between CPPR = 400 and CPPR = 800. 
Thus, our specific choice of CPPR = 400 is assured for the estimate of $F_{ch}$,
and will be adopted in all the following simulations in this paper. 
Note that we also perform a simulation of a collision with a flat target
and confirm that our results ($F_{ch}$) are in good agreement with the previous work \citep{jmm15} (not shown in this paper).

\begin{figure}
\figurenum{4}
\plotone{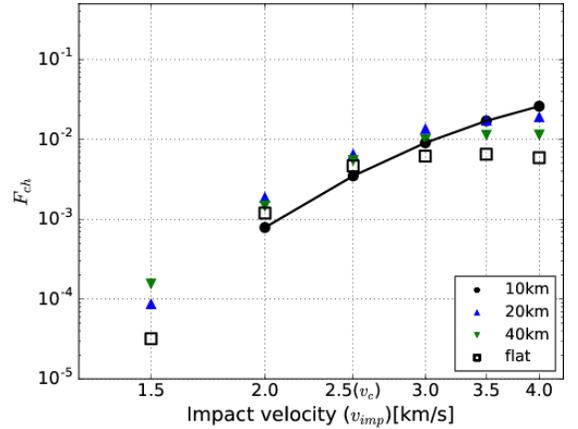}
\caption{The value of $F_{ch}$ obtained from different impact velocities and target's diameters. 
The results are gained from collisions of a 10 km, 20 km, 40 km and flat (represents protoplanet) sized target (see legends) with a 10 km sized impactor. 
Impact velocity of collisions varies from 1.5 km s$^{-1}$ to 4.0 km s$^{-1}$ with the interval of 0.5 km s$^{-1}$.
The solid line denotes the results for the case of 10 km sized targets.
\label{fig:Fch}}
\end{figure}

\begin{figure}
\figurenum{5}
\plotone{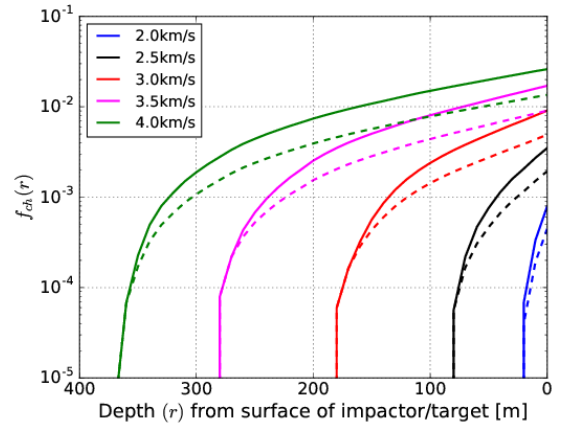}
\caption{The resultant value of $f_{ch}(r)$ from an impactor (10 km) and a target (10 km) as a function of depth ($r$) from their surface (see equation (\ref{eq:f_ch_r})).
The dashed lines denote $f_{ch}(r)$ from the impactor and the solid lines denote the summation of $f_{ch}(r)$ both from the impactor and from the target.
The results for different impact velocities are denoted by different colors (see legends).
\label{fig:10kmDepth}}
\end{figure}

\subsection{Dependence on impact velocity for 10km sized target} \label{sec:10km}

We examine how $F_{ch}$ varies as a function of the impact velocity for collisions between the same sized planetesimals. 

Figure \ref{fig:Fch} shows the resultant behavior for collisions with the 10 km sized target planetesimals (see the solid line). 
We find that as $v_{imp}$ increases $F_{ch}$ also increases.
This occurs simply because a high impact velocity generates large impact energy, which creates more chondrules.
Interestingly, even if $v_{imp}$ = 2.0 km s$^{-1}$, 
which is lower than the threshold velocity adopted in the previous work ($v_c$= 2.5 km s$^{-1}$), 
chondrules could form for the same sized planetesimal collisions. 
Since the impact velocity of $v_{imp}$ = 2.0 km s$^{-1}$ can be achieved at the early stage of planetary accretion, 
our results suggest that when planetesimal-planetesimal collisions are considered,
impact jetting can start forming chondrules at an earlier time than what the previous studies indicate \citep{jmm15,hwmo16,htm16}.
This finding is important because the meteoritic data suggest that the formation time of the oldest chondrules overlaps with that of CAIs \citep{cbk12}.
In other words, impact jetting may be able to produce such chondrules via planetesimal-planetesimal collisions, if planetesimals and CAIs formed contemporary
(see \S\ref{sec:diff} for a more discussion).
Note that the value of $F_{ch}$ is small (0.1 \%) for this case, 
so that the mass budget of chondrules may not be affected by the collisions significantly.
We also find that when the impact velocity is slower than 1.5 km s$^{-1}$, 
the resultant chondrule formation would be negligible since $F_{ch}$ is much less than $10^{-3}$ \%.

Figure \ref{fig:10kmDepth} shows the cumulative mass fraction of chondrules ($f_{ch}(r)$) 
as a function of depth ($r$) from the surface of an impactor and a target (see equation (\ref{eq:f_ch_r})). 
The contribution from the impactor is denoted by the dashed line, 
and the total ejected mass both from the impactor and form the target is by the solid line.
In other words, the contribution only from the target is represented by the difference between the solid and the dashed lines.
It clearly shows that progenitor of chondrules can be ejected at deeper regions from the surface when collisions occur with higher impact velocities.
Our results also show that the value of $f_{ch}(r)$ increases at the same depth for higher impact velocities.
These explain the results shown in Figure \ref{fig:Fch}: $F_{ch}$ is an increasing function of the impact velocity.
Furthermore, we find that the deepest region that can eject progenitor of chondrules is located only at about 400 m from the surface 
even for the case of $v_{imp}$ = 4.0 km s$^{-1}$. 
It is important that half of the chondrule mass ($F_{ch}$) arises from a very thin surface layer, 
which is within 100 m from the surface.
In addition, our results show that the chondrule masses formed by impact jetting from the impactor and the target are the almost the same.
Planetesimals with the diameter of 10 km hardly experience differentiation especially at their surface. 
Therefore, produced ejecta from such collisions can surely become chondrules which can be compatible with the ones found in chondrites.

\begin{figure}
\figurenum{6}
\plotone{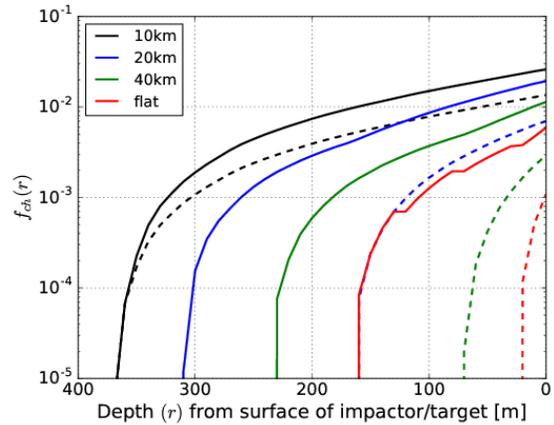}
\caption{The resultant value of $f_{ch}(r)$ from an impactor and a target as a function of depth ($r$) from their surface (as in Figure \ref{fig:10kmDepth}).
The size of a target varies from 10km to flat (see legends) with $v_{imp}$ = 4.0 km s$^{-1}$.
\label{fig:depth}}
\end{figure}

\subsection{Dependence on target sizes with $v_{imp} = 4.0$ km s$^{-1}$} \label{sec:v40}

We explore how the diameter of target planetesimals affects impact jetting and the resultant chondrule formation. 
In this section, we adopt the impact velocity of 4.0 km s$^{-1}$ for all the runs and change only the diameter of target. 
In Section \ref{sec:all}, we will discuss  the results in which both the impact velocity and the target's size vary.

Figure \ref{fig:Fch} shows that the total mass of chondrules becomes the largest when a 10 km sized planetesimal serves as a target.
As the diameter of target planetesimals increases, the produced mass of chondrule becomes smaller. 
Our simulations nonetheless show that the resultant values of $F_{ch}$ are comparable for a wide range of the target size.
We can therefore conclude that the results obtained from flat targets can be used as a reference value 
for both planetesimal-planetesimal and protoplanet-planetesimal collisions.
More specifically, the chondrule mass estimated by the previous studies would still hold 
even for planetesimal-planetesimal collisions \citep{jmm15,hwmo16,htm16}.

We also plot the cumulative chondrule mass ($f_{ch}(r)$) from the target and the impactor 
as a function of depth ($r$) from their surface in Figure \ref{fig:depth} (as done in Figure \ref{fig:10kmDepth}). 
For the case of collisions with a 10 km sized target, progenitor of chondrules is ejected from the deepest region ($r \simeq 400$ m).
Equivalently, the largest amount of chondrules is produced.
We find that $f_{ch}(r)$ decreases at all the depth with increasing the size of targets.
This indicates that the original locations of chondrules are the key to evaluating the value of $f_{ch}(r)$ (and hence $F_{ch}$).
It is interesting that progenitor of chondrules emerges from a deeper region from targets than impactors
when targets are bigger than impactors.
Thus, our results suggest that the composition of targets is more important for determining that of chondrules more accurately
when there is a size difference between targets and impactors.

\subsection{Dependence on impact velocities and target sizes} \label{sec:all}

We are now in a position to change both the impact velocity and the target size to examine how these two parameters affect chondrule formation. 

Figure \ref{fig:Fch} shows the corresponding results. 
We find that the behaviors of $F_{ch}$ are very complicated.
For instance, we consider collisions with larger target planetesimals ($\ge$ 40 km in diameter, see the green triangles and the open squares).
When the impact velocity is slower than 3.5 km s$^{-1}$, the value of $F_{ch}$ increases monotonically.
Once the impact velocity exceeds 3.5 km s$^{-1}$, however,
the value becomes smaller.
It should be pointed out that such complicated behaviors are also observed in the previous work \citep{jbm14},
while our numerical setups are not entirely the same as the work.

Based on this parameter study, we may be able to draw the following two conclusions.
The first one is that when the impact velocity is smaller than $< 2.0$ km s$^{-1}$, 
the resultant value of $F_{ch}$ is around  0.1 $\%$.
Consequently, such collisions may be important to regulate the formation timing of chondrules, but not their total abundnace.
The second one is that while the dependence of $F_{ch}$ on impact velocities and the target sizes is not so simple,
the condition adopted by the previous work can be applicable for various kinds of collisions that can occur during planetary accretion:
chondrule formation can take place at planetesimal collisions with the impact velocity of $v_{imp} \simeq 2.5$ km s$^{-1}$ or higher, 
and the resultant abundance of chondrules formed by a collision is $F_{ch}m_{pl}$ with $F_{ch}$ of 0.01.

\section{Discussion}

As demonstrated above, planetesimal collisions and the resultant impact jetting can play an important role in forming progenitor of chondrules.
Our results, however, are obtained under the assumptions with which the numerical simulations are idealized. 
We here examine how such assumptions are reasonable and briefly comment on to what extent our results can be affected by the assumptions.
We also discuss a potential effect of differentiation that may occur for colliding planetesimals, 
and the fate of ejecta that can eventually become chondrules.
Finally, we compare our results with chondrules actually found in chondrites.

\subsection{Numerical setups} \label{sec:num}

We here discuss the assumptions that are used in our numerical simulations.

First, we assume that the bulk compositions of an impactor and a target are the same in our simulations; porous dunite.
It is well known that ordinary chondrites have a porosity of $< 10\%$ \citep{cbm08}.
The value of porosity adopted in our simulations, which is $1\%$, is in such a range.
The bulk composition of planetesimals used in this paper, dunite, is similar to that of ordinary chondrites \citep{ss15}.
Thus, our assumption that colliding, undifferentiated planetesimals would be represented by porous dunite may be reasonable,
and hence our results would be useful for producing chondrules found in ordinary chondrites.
Note that the properties of melt ejecta that can finally be transformed to chondrules
would be determined both by the porosity and by the composition of impactors and/or targets \citep{wco08}.

Second, we fix the initial temperature of planetesimals at 300 K in all the simulations.
If this value would be decreased down to $\sim 150$ K, 
(which is more appropriate for the gas temperature at 2$-$3 au in the solar nebula), 
the resultant mass of melting ejecta would be smaller. 
On the contrary, if the surface of planetesimals has a much higher temperature ($>$ 600K) following the formation and growth of planetesimals,
then the produced chondrule mass might increase \citep[e.g.,][]{ss12}. 
A more comprehensive parameter study would be needed to fully develop a quantitative argument.

Third, only head-on collisions are examined in this paper.
This arises because we have made use of the iSALE-2D code.
Such collisions would probably produce a lower amount of chondrules, compared with other types of collisions.
Thus, our results are very likely to provide a conservative estimate in the resultant abundance of chondrules.
In our future work, we will use the iSALE-3D code
to explore the three-dimensional effect on planetesimal collisions and compute the amount of chondrules formed by the collisions.

\subsection{Effect of differentiation} \label{sec:diff}

We discuss a possibility about the occurrence of differentiation in planetesimals and its effect on chondrule formation via impact jetting.

Differentiation is indeed an important process for the impact jetting scenario to produce chondrules.
This is because once planetesimals experience differentiation,
the composition of colliding planetesimals may not be entirely the same as that of the currently sampled chondrules.
As a result, it may be difficult to generate chondrules from ejected materials.
In addition, while our results suggest that chondrules formed via planetesimal-planetesimal collisions might be able to generate 
the oldest chondrules (see \S\ref{sec:10km}), this suggestion would be valid only if colliding planetesimals did not experience differentiation at that time.
Thus, differentiation can affect both the composition of chondrules and their formation timing.

Based on thermal modelling of planetesimals, 
the onset of differentiation in planetesimals is regulated predominantly by their formation timing \citep[e.g.,][]{gtbs14,wniy14}.
When planetesimals formed at a few Myrs after CAI formation, 
which occurred 4567 Myr ago from now \citep{cbk12},
most planetesimals are hard to differentiate.
Differentiation might occur for planetesimals that are larger than 20 km in diameter, 
when they formed shortly after the formation of CAIs.

In our model, the formation of planetesimals has to occur prior to the formation of chondrules,
since planetesimal collisions are the mechanism to generate chondrules.
In other words, the impact jetting scenario can become a crucial process for chondrule formation
only if the formation of planetesimals took place in the solar nebula well after the CAI formation
and these planetesimals underwent impact jetting.
Given that it is still poorly known how to form planetesimals \citep{jbmt14},
it is unlikely that a definitive statement can be made here.
It may nonetheless be possible to suggest that if planetesimal formation occurred continuously in the solar nebula,
impact jetting can contribute to chondrule formation for a long time.
In fact, chondrule formation continued 3-5 Myr after CAI formation started \citep{cbk12,bcb15}.
Furthermore, in order for impact jetting to generate the oldest chondrules, colliding planetesimals should have been small enough 
($\sim$ 10 km in diameter) to avoid differentiation. 
Taking into account that at least about $10^5$ yrs are needed to form protoplanets once a large number of planetesimals are present \citep{jmm15,hwmo16}, 
planetesimal-planetesimal collisions would have more chance to generate the oldest chondrules via impact jetting (than protoplanet-planetesimal collisions).

It should be noted that the above argument has been made based on the assumption that when colliding bodies are fully differentiated, 
they cannot yield jetted materials of an undifferentiated composition. 
As pointed out by \citet{jmm15}, however, differentiated bodies may be able to have surface layers that contain yet-to-be differentiated materials as long as the bodies are still growing \citep[e.g.,][]{we13}. 
Since our results show that the progenitor of chondrules arises from thin ($\sim$ 100 m) surface layers (see Figures \ref{fig:10kmDepth} and \ref{fig:depth}), 
there may still be a possibility that largely differentiated, growing bodies (such as massive planetesimals and protoplanets) can generate  chondrules via impact jetting.
A more comprehensive work would be needed to fully address these issues, wherein thermal modeling of massive planetesimals 
(and/or protoplanets) and the impact jetting process are examined simultaneously. 

\subsection{The fate of melting ejecta} \label{sec:fate}

While we have so far computed the mass fraction ($F_{ch}$) of ejecta that experience a high temperature ($T_m$) and have a high ejection velocity ($v_{imp}$),
we have not discussed the fate of such ejecta yet.
Here, we attempt to explore how the ejecta can eventually become chondrules and will be accreted by parent bodies of (ordinary) chondrites.

As suggested by \citet{jmm15}, chondrules can be formed from the resultant melts via impact jetting.
For this case, the typical size of chondrules could be estimated from the surface tension and the velocity of ejecta \citep{mv91, jm14, jmm15}.
We here apply a formula \citep[see equation (6) in][]{jm14} to our results
under the assumption that the surface tension of dunite is the same as that of silica.
We find that the diameter of chondrules generated from the impact jetting would be around 0.4 mm.
The high impact velocity results in the high velocity of ejecta,
which leads to the production of smaller chondrules.
Our results nonetheless show that the diameter of chondrules shrinks down to only 0.2 mm for the highest impact velocity case.
Based on the above simulations, we have confirmed that
the velocity of ejecta does not depend on the size of target planetesimals very much.
As a result, the diameter of chondrules is not so sensitive to the impact velocities and the size of targets.

Could the resultant droplets of chondrules return or be accreted onto planetesimals that can serve as parent bodies of chondrites? 
\citet{htm16} discussed this based on the idea of pebble accretion in a magnetized disk. 
Small particles including chondrules can be accreted onto massive planetesimals via pebble accretion \citep[e.g.,][]{jmlb15}. 
They utilize the estimate of the magnetic field strength that is derived from chondrules in Semarkona ordinary chondrite \citep{fwl14}.
They found that in order for planetesimals to accrete chondrules and to become parent bodies of chondrites,
the solar nebula should be $< 5$ times more massive than the minimum mass solar nebula \citep{h81} and 
the planetesimals are $< 500$ km in diameter ($\sim 10^{21}$ kg). 
\citet{mohw16} undertook a follow-up study in which pebble accretion of chondrules in a gas disk is investigated carefully.
They consider a situation that a protoplanet, planetesimals and chondrules all coexist. 
They have found that although about half of produced chondrules are accreted onto the protoplanet, 
ten percent of chondrules are accreted onto planetesimals and can make a thin chondrule-rich layer near their surface.
If such planetesimals can act as parent bodies of chondrites, 
it may be possible to create chondrites with the high abundance of chondrules;
chondrites would be fragments originating from the surface region of the planetesimals.

In addition to the ejecta that can eventually become chondrules, it would also be important to discuss the fate of unmelted ejecta that cannot be transformed to chondrules. 
This is because planetesimal-planetesimal collisions can potentially generate a considerable number of such unmelted broken rocks (see Figures \ref{fig1} and \ref{fig2}).
\footnote{As shown by \citet{jmm15}, the overall results of protoplanet-planetesimal collisions are exclusively melted materials -  the progenitor of chondrules.} 
The presence of such broken rocks may become problematic in reproducing the high abundance of chondrules in ordinary chondrites;
a good mixture of chondrules and broken rocks may end up with difficulty generating purely chondritic meteorites. 
One of the possible processes to filter out the broken rocks from a swarm of chondrules may arise from a size difference between chondrules and the broken rocks. 
As discussed above, small-sized particles like chondrules will be accreted onto massive bodies via pebble accretion. 
If the broken rocks would be much larger than or much smaller than chondrules, then the accretion timescale of such rocks onto massive planetesimals/protoplanets would be different from that of chondrules. 
As a result, it may be possible to selectively remove unmelted broken rocks from a collection of chondrules, and hence to create purely chondritic meteorites. 
An intriguing conjecture can be developed from this argument: since the particle size is the main quantity to trigger this filtering process, some chondrites may possess both chondrules and unmelted rocks with the size of $\sim$ 1mm. 
It is interesting that some (breccia) chondrites indeed contain the unmelted and lightly shocked, chondrule-sized materials  \citep[e.g.,][]{sks91}. 
The detail will remain to be explored for future work.

Thus, the coupling of impact jetting with pebble accretion of chondrules is promising 
to obtain new insights about the formation of chondrules, chondrites, and up to planets in the solar system.

\subsection{Comparison with chondrules found in chondrites} \label{sec:chond}

It is interesting to compare our results with the actually found chondrules.

As discussed above, our results suggest that chondrules are about 0.4 mm in diameter.
On the other hand, the diameter of chondrules in ordinary chondrites is estimated as 0.3 - 0.6 mm \citep{sk05,s07,fwe15}.
Thus, our results are in good agreement with the actual measurements of chondrules in ordinary chondrites. 

While we have focused mainly on chondrules in ordinary chondrites in above sections,
chondrules are also found in carbonaceous chondrites.
Since most chondrules are composed mainly of silicate, our results may be applicable for chondrules in CB chondrites (one type of carbonaceous chondrites) as well.
In fact, it is suggested that chondrules in CB chondrites would form via collisional processes \citep{kacm05,bcb15}.
Note that the bulk composition of CB chondrites are quite different from that of ordinary chondrites,
so that parent bodies of these two chondrites should be largely different with each other.
The diameter of chondrules in CBb chondrites (subgroup of CB chondrites) is $\sim$ 0.5 mm \citep{sk05,s07},
which is comparable to the size of our results.
Thus, planetesimal collisions and the subsequent impact jetting may be a crucial process to account for the properties of chondrules found in chondrites.

It should be noted that chondrites contain various sizes of chondrules and their texture also varies \citep{sk05,s07}.
For instance, the diameter of chondrules in CBa chondrites (another subgroup of CB chondrites) is measured as $\sim$ 5 mm \citep{sk05,s07,bcb15},
which is much larger than what our results predict.
A more detailed analysis would be needed to fully understand the properties of currently sampled chondrules 
such as their size distribution, their mineralogy and their texture. 

\section{Conclusions}

We perform numerical simulations of planetesimal collisions using the shock physics code,
and compute the mass of ejecta that can eventually become chondrules.
We especially focus on undifferentiated planetesimal-planetesimal collisions.
This is because if target planetesimals are not fully differentiated, 
the resultant ejecta could be transformed to chondrules as long as their bulk compositions are similar to that of chondrules.

We examine a relationship between the mass fraction ($F_{ch}$) of progenitor of chondrules and the parameters of collisions; the impact velocity and the size of targets.
The size of target planetesimals covers the range from 10 km to 50 km in diameter,
and the impact velocities ($v_{imp}$) are from 1.0 km s$^{-1}$ to 4.0 km s $^{-1}$.
We find that $F_{ch}$ would be around 10$^{-2}$ when $v_{imp}$ is equal to or larger than 2.5 km s$^{-1}$.
The progenitors of chondrules come from a thin surface layer of impactors and/or targets ($400$ m at the most).
Our results show that $F_{ch}$ is correlated with the depth from the surface of impactor and target planetesimals
at which progenitors of chondrules are ejected.
Planetesimal-planetesimal collisions can also produce chondrules as protoplanet-planetesimal collisions \citep{jmm15}.
Additionally, the former collisions can accelerate the formation time of chondrules 
while their contribution to the produced chondrule mass may not be so significant \citep{hwmo16}.
As a result, the previous studies of impact jetting \citep{jmm15,hwmo16,htm16} may overestimate when chondrule formation begins
due to the assumption that $v_c$ and $F_{ch}$ are constant for any collisions.

In a subsequent paper, we will perform full N-body simulations coupled with the results of this study,
wherein both $v_c$ and $F_{ch}$ depend on the properties of collisions,
and compute formation timing and the resultant mass of chondrules more realistically. 
 
\acknowledgments
We gratefully acknowledge the developers of iSALE-2D, including Gareth Collins, Kai Wunnemann, Dirk Elbeshausen, Boris Ivanov and Jay Melosh.
Numerical computations were carried out on the PC cluster at Center for Computational Astrophysics, National Astronomical Observatory of Japan.
We thank the referee, H. J. Melosh for helpful suggestions and comments.
S. W. thank Alexander N. Krot for helpful comments and discussions on chondrules and also thank Brandon C. Johnson and 
Kosuke Kurosawa for kindly comments on simulations.
The part of this research was carried out at JPL/Caltech under a contract with NASA. Y. H. is supported by JPL/Caltech.

\end{document}